\documentclass[12pt]{article}
 \input{epsf}

 \usepackage{epsfig}

\begin{document}

 \title{The stability of vacua in two-dimensional gauge theory}
 \author{Victor Kac \thanks{kac@math.mit.edu} 
  \\ Department of Mathematics \\
Jan Troost \thanks{ troost@mit.edu ;  Preprint  MIT-CTP-3039. } \\ 
    Center for Theoretical Physics \\ \hspace{3cm} \\ MIT \\
    77 Mass Ave \\ Cambridge, MA 02139 \\ USA 
   }

\maketitle
 \abstract{We discuss the stability of vacua in two-dimensional
gauge theory for any simple, simply connected gauge group. Making use
of the representation of a vacuum in terms of a Wilson line
at infinity, we determine which vacua are stable against pair
production of heavy matter in the adjoint of the gauge group. By
calculating correlators of Wilson loops, we reduce the 
problem to a problem in representation theory of Lie groups,
that we solve in full generality.} 

 \setcounter{equation}{0}
 \section{Introduction}

Two-dimensional gauge theory has often been used as a toy model
for four-dimensional gauge theory. Although there are no propagating
degrees of freedom in two dimensions, two-dimensional gauge theory
(supplemented with appropriate matter)
can mimick some of the most intriguing features of QCD in 4 dimensions,
like confinement, a fermion condensate, a non-trivial vacuum structure and
stringy behavior at large $N$
(see e.g. \cite{Getall}).

In this letter we study two-dimensional gauge theory for a simple,
simply connected gauge group $G$, with massive
 matter transforming in the adjoint.
The effective gauge group is then $G/Z$, where $Z$ is the center of the
group $G$.
The non-trivial vacuum structure can be gathered from homotopical
arguments. Just as $\Pi_3 (SU(3)) = {\bf Z}$ leads to $\theta$-vacua in
four dimensions, so $\Pi_1 (G/Z) = Z$ makes  ${\#} Z$ $\theta$-vacua in
two dimensions plausible \cite{IS}. In this letter,
 we want to study the existence and
stablity of these vacua in more detail --  the dynamics involved
in the stability of the vacua cannot be read from the topological reasoning.

There are no propagating degrees of freedom for a gauge field in two
 dimensions.
Therefore, to study dynamical aspects of the vacua,
 we introduce massive adjoint matter (that can mimick aspects of the dynamics
of transverse gluons in higher dimensions (see e.g. \cite{DK})).
 We will study the stability of vacua against
pair creation, by representing
them as Wilson lines at infinity \cite{W1}. We have a discrete choice for the 
representation at infinity, since the gauge group is a simple,
simply connected Lie group.
 In the limit where the adjoint particles are heavy, we can evaluate
their interaction by computing the correlation function of the corresponding
Wilson line with the Wilson line at infinity.
We will study whether the strings  stretching between our heavy
probes have positive or negative tension. If a negative string tension
is generated, this implies that the choice for representation
for the vacuum at infinity is unstable against pair creation \cite{W1}
\cite{PSZ2}. 
Moreover, as we will see, we also have to check the stability of the
vacuum against the creation of more than one particle anti-particle pair
 to find agreement with topological
arguments \cite{PSZ1}.

These phycical problems reduce to neat problems in representation theory
of Lie groups. We solve these problems in all generality, completing the
analysis performed in \cite{PSZ1}.  

\section{Vacua as Wilson lines at infinity}

In two-dimensional gauge theory, we can represent a choice of vacuum
by a choice for the representation for a Wilson line at infinity
\cite{W1}. This is perhaps most familiar in two-dimensional abelian
gauge theory, where charges at infinity mimick the existence of
$ \theta $-vacua that are associated to the non-trivial homotopy group
$\Pi_1 (U(1))  = \mbox{{\bf Z}}   $. This is easily demonstrated since, for 
a $U(1)$ gauge theory with Wilson loop of charge $\frac{\theta}{2 \pi}$
at infinity,
we have:
\begin{eqnarray}
< W_{\infty} (\theta) > &=&
\int dA \, e^{\int F \wedge \ast F} e^{\frac{\theta}{2 \pi} \oint_{\infty} A} \nonumber \\
&= & \int dA \, e^{\int F \wedge \ast F + \frac{\theta}{2 \pi} \int F}.
\end{eqnarray}
Therefore, correlation functions evaluated with the insertion of 
a Wilson loop at infinity,
or in a $\theta$-vacuum, are identical.

For our non-abelian gauge theory with simple gauge group $G$, there is 
a similar representation, but the choice of Wilson loop is the discrete
choice of irreducible representation in which we evaluate the trace
of the path-ordered exponential \cite{W1}.

\section{Adjoint matter and stability of vacua}

We introduce adjoint matter into the system, to obtain non-trivial
dynamics that may be taken to mimick the behavior of the transverse gluons in
higher dimensional Yang-Mills theories. We will take these particles
to be heavy, such that we can measure their interaction 
by evaluating the expectation value of the associated Wilson loop, 
in the adjoint of the gauge group \cite{Wilson}.

The Wilson loop calculation is straightforward, using standard 
techniques in two-dimensional gauge theory 
\cite{M} \cite{R} \cite{W2}  \cite{BT} \cite{PSZ1}. 
To calculate the correlator of the two Wilson loops,
we glue a Wilson line at infinity in an arbitrary representation $R(\mu)$
with highest weight $\mu$,
into the propagator on the cylinder, and at the end of the cylinder,
we put another Wilson loop in the adjoint $ R(\theta)$,
 representing the particle anti-particle pair,
and then we close the two-dimensional plane using the propagator
on the disc. \footnote{These operations are systematically explained in \cite{BT}. We heavily 
make use of the invariance of 2d Yang-Mills theory under area preserving
diffeomorphisms.}
We denote by $g$ the Yang-Mills coupling, by $A_2$ the area enclosed by
the inner  Wilson loop in the adjoint and by $A_1$ the area of the rest of the 
plane. $C(\lambda)$ is the quadratic Casimir evaluated in the representation
with highest weight $\lambda$. We have then:
\footnote{We abuse notation by indicating a representation
by its highest weight, for brevity. For notations, see also appendix
\ref{theorem}.}
\begin{eqnarray}
<W_{\infty} (\mu) W (\theta) > &=& \sum_{\lambda,\lambda'} \int_G dg \,  dg'
\chi_{\mu} (g) \chi_{\lambda} (g^{-1}) e^{-\frac{g^2}{2} C(\lambda) A_1} \nonumber \\
& & \chi_{\lambda} (g')
\chi_{\theta}(g') \chi_{\lambda'} ((g')^{-1}) e^{-\frac{g^2}{2} C(\lambda') A_2} d(\lambda') \nonumber \\
&=& \sum_{       \lambda'}           \int_G dg'
 e^{-\frac{g^2}{2} C(\mu) A_1} \chi_{\mu} (g')
\chi_{\theta}(g') \nonumber \\ & & 
\chi_{\lambda'} ((g')^{-1}) e^{-\frac{g^2}{2} C(\lambda') A_2} d(\lambda') \nonumber \\
&=& \sum_{ \nu_i \subset \mu \otimes \theta}  e^{-\frac{g^2}{2} C(\mu) A_1} e^{-\frac{g^2}{2} C(\nu_i) A_2} d(\nu_i) \nonumber \\
&=& \sum_{\nu_i \subset \mu \otimes \theta}d(\nu_i)  e^{-\frac{g^2}{2} C(\mu) (A_1+A_2)} 
e^{-\frac{g^2}{2} (C(\nu_i)-C(\mu)) A_2}. \nonumber
\end{eqnarray}
After we factored out the contribution
 corresponding to the exterior Wilson loop
at infinity, we  can interpret the exponent
 $C(\nu_i)-C(\mu)$ as the string tension for the string
stretching between the adjoint matter.\footnote{ This intuitive picture was
corroborated in \cite{PSZ2} by
 the corresponding detailed Hamiltonian analysis.}
We need $C(\nu_i) \ge C(\mu)$ for all
$\nu_i \subset  \mu \otimes \theta$ \footnote{The notation $\subset$ 
indicates that the irreducible representation $\nu_i$ is part
 of the decomposition of the tensor product 
in irreps.} to have a vacuum that is stable against
pair creation. In the next section, we enumerate the
representations $R(\mu)$ satisfying this condition, and we prove that our
list is complete. 

But this is not the end of the story. In fact, we also want to consider
whether
the vacuum is stable against the creation of more than one particle pair
 \cite{PSZ1}. After a calculation similar to the one we performed in 
detail above, this boils down to the following 
extra condition:
$ C(\nu_i) \ge C(\mu) $ for all $\nu_i \subset  \mu \otimes (\theta)^n$,
where $n$ is the (arbitrary) 
number of matter pairs that are created in the process
of screening the trial vacuum at infinity.

 Following \cite{PSZ1} we call the
vacua that are stable against the appearance of a single pair of adjoint
particles metastable. The vacua satisfying the more stringent 
condition are the stable vacua. 
In the next section, by solving these problems in representation theory,
we give a complete list of metastable and stable vacua, and prove the list
\cite{PSZ1} in 
all generality.

\section{Stable and metastable vacua}
\subsection{Metastable vacua}
{\bf Problem:} Find all representations $\mu$ such that
\begin{eqnarray}
C(\nu_i) \ge C(\mu) \quad \mbox{for all} \quad \nu_i \subset  \mu
 \otimes \theta.
\label{metastable}
\end{eqnarray}
We call these representations and corresponding highest weights
 $\mu$ metastable.
\newline
{\bf Answer.} The metastable weights for gauge group $G$
are given by the level 1 weights of the corresponding affine (non-twisted)
Kac-Moody group $\hat{G}$. \footnote{For nomenclature,
see also \cite{K} \cite{GO1}.}
 In appendix \ref{theorem} and
\ref{Dynkin} we give our
conventions and in appendix \ref{Dynkin} we list
the level 1 weights. To prove this statement, we first introduce the notion
of a 'bad root'. We call a positive root $\alpha$ bad for a representation
$\mu$ (with labels $m_i=\mu.\check{\alpha_i}$) if $\alpha-(m_i+1) \alpha_i$ is not a root or $0$ for all $i$,
and $(\mu+\rho).\check{\alpha}>1$. Here and further $\alpha_1, \dots, 
\alpha_r$ are the simple roots and $\check{\alpha}=\frac{2 \alpha}{\alpha^2}$

We claim that $\mu$ is metastable if and only if there are no positive bad
roots for $\mu$. Indeed, 
all highest weights $\nu_i$ in the decomposition are
of the form $\mu -\alpha$, where $\alpha$ is a root or $0$. The inequality
in (\ref{metastable}) becomes then:
 \begin{eqnarray}
|\nu_i+\rho|^2 &\ge& |\mu+\rho|^2 \Leftrightarrow \nonumber \\
  (\rho+\mu). \check{\alpha} &\le& 1. \label{bad}
\end{eqnarray}
If $\alpha \le 0$, this inequality always holds. If $\alpha>0$, then 
by the theorem in appendix \ref{theorem}, $\mu-\alpha=\nu_j$
for some $j$  if 
and only if $\alpha-(m_i+1) \alpha_i$ is not a root or $0$ (where
$m_i$ are the labels of $\mu$). In that case the 
inequality (\ref{bad}) must hold.
 This proves our claim.
We are now prepared to prove the answer in two steps: 
\newline
Step 1: All weights satisfying (\ref{metastable}) are fundamental or $0$.
\newline
 a) No weight $\mu$ has Dynkin label 2 or more. Indeed, 
it is clear that $\alpha_i$
would be a bad root for $\mu$.
\newline
b) No two Dynkin labels
equal to 1 are allowed. Suppose $m_i=1=m_j \quad (i \neq j)$.
Then the root
 $\alpha=\alpha_i + \alpha_{i+1} \dots + \alpha_{j-1} + \alpha_j$,
that connects dot $i$ with dot $j$ in the Dynkin diagram of $G$,
is a bad root.
\footnote{The indices do not necessarily differ by one. See appendix
 \ref{Dynkin} for our conventions for labeling simple roots.}
 We conclude that only
fundamental weights can satisfy condition (\ref{metastable}).
\newline 
Step 2:
 Firstly, observe that $\mu=\theta$, the highest root, is not a metastable weight (since $0 \subset \theta \otimes \theta$).
Secondly, if $\mu$ is not metastable for a subdiagram of the Dynkin diagram
of $G$, then $\mu$ is not metastable for $G$ (due to the
above claim). These two observations 
along with Table 1 and Figure 1 show that
all non level 1 fundamental weights $\lambda_i$ are not metastable,
except for $\lambda_2$ for $F_4$. The latter is not metastable since the root
$\alpha_1+3\alpha_2+2 \alpha_3+\alpha_4$ is bad for $\lambda_2$. That
 the level 1
 weights do satisfy the criterium (\ref{metastable}),
can be checked on a case by case
basis by going over all positive roots (listed e.g. in \cite{OV})
and verifying that all of them are good. (For miniscule weights, in particular
in all simply-laced cases, this also follows from the geometric argument
given in section \ref{substable}.)
\subsection{Stable vacua}
\label{substable}
{\bf Problem:} Find all representations $\mu$ such that
\begin{eqnarray}
C(\nu_i) \ge C(\mu) \quad \mbox{for all} \quad \nu_i \subset  \mu \otimes \theta^n
\quad
\mbox{for any} \quad n
\label{stable}
\end{eqnarray}
{\bf Answer.}
The stable vacua are given by the miniscule weights of $G$.\footnote{For
nomenclature see \cite{B} \cite{GO2} and appendix \ref{Dynkin}. There are
$\# Z$ miniscule weights.}
 We prove this in two steps.
\newline Step 1: We rule out the level 1 non-miniscule weights.
We distinguish the following chains of representations:
\newline
For $G_2$:
\footnote{These decompositions were obtained using the software \cite{soft};
 they can also be obtained using the theorem in 
appendix \ref{theorem}.}
\begin{eqnarray}
\lambda_1 \otimes \lambda_2 \supset 2 \lambda_1 \nonumber \\
 2 \lambda_1 \otimes \lambda_2 \supset \lambda_2 \nonumber \\
\lambda_2 \otimes  \lambda_2 \supset 0
\end{eqnarray}
For $F_4$:
\begin{eqnarray}
\lambda_1 \otimes \lambda_4 \supset  \lambda_2 \nonumber \\
  \lambda_2 \otimes \lambda_4 \supset 2 \lambda_1 \nonumber \\
2 \lambda_1 \otimes  \lambda_4 \supset \lambda_4 \nonumber \\
 \lambda_4 \otimes  \lambda_4 \supset 0
\end{eqnarray}
\newline
For $B_r$:
From the tables in \cite{OV}, we find that, starting with $\lambda_1$,
tensoring with the adjoint $n$ times contains $\lambda_{2n+1}$,
untill we hit the end of the Dynkin diagram. There, regardless of whether
$r$ is odd or even, we find that the representation $2 \lambda_r$ is contained
in the decomposition of $\lambda_1 \otimes \theta^{[r/2]}$. Tensoring 
$2 \lambda_r$
 with the adjoint brings us back over the complementary fundamental
weights to $\lambda_2$. Tensoring once more with the adjoint, which is
$\lambda_2$, give us a trivial representation.
\newline
For $C_r$:
Using the tables in \cite{OV} or the 
theorem in appendix \ref{theorem}, we find that
generically $\lambda_k \otimes 2 \lambda_1 \supset \lambda_1 + \lambda_{k+1}$.
Similarly  $ (\lambda_1 + \lambda_{k+1}) \otimes 2 \lambda_1 \supset  \lambda_{k+2}$. Therefore all fundamental weights with even index are linked by chains
of representations,
as are all fundamental weights with odd index. Note also that
 $\lambda_2 \otimes 2 \lambda_1 \supset 2 \lambda_1 $. It is clear then
that all odd fundamental weights are screened to the miniscule weight $\lambda_1$, and
all even to the trivial weight $0$, since these have smaller Casimir then
all other fundamental weights. (This will also be clear from the geometrical
proof in step 2).
\newline Step 2: We proof geometrically that all miniscule weights satisfy condition (\ref{stable}).
 The fundamental alcove $A$ is the part of the fundamental chamber
$C$ that satisfies $\mu.\check{\alpha_i} \ge 0$ and 
$\mu.\check{\theta}' \le 1$,
where $\theta'$ is the highest short root.
It intersects with the dominant weights only by the miniscule weights 
\cite{B}.  Clearly it is sufficient to prove that if $\mu+ \alpha$
lies in the fundamental chamber $C$, for $\mu$ miniscule 
and $\alpha$ a nonzero in the root
lattice $Q$,  then:
\begin{eqnarray}
|\mu+\alpha+\rho|^2 &>& |\mu + \rho|^2 \quad \mbox{or} \nonumber \\
(2 \mu + 2 \rho+ \alpha) . \alpha &>& 0.
\end{eqnarray}
Now, $C$ is covered by alcoves $\nu+A$, where $\nu \in Q \cap C$.
Since $\mu \in A$ and $\mu + \alpha \in C$, we conclude that $\alpha \in C$.
Since the inverse of the Cartan matrix has only positive entries, we conclude
that $\alpha$ is a linear combination of simple roots with non-negative
coefficients. Hence $\mu . \alpha \ge 0$,  $(\mu+\alpha) . \alpha \ge 0$ ,
and $\rho . \alpha > 0$. QED.

\section{Conclusion}
We proved that the classification of stable vacua in two-dimensional
gauge theory with simple, simply connected gauge group $G$ and heavy 
matter in the adjoint, in
an algebraic fashion based on Wilson loop computations, agrees with
the naive homotopical arguments for the classification of vacua.
The stable vacua correspond to the miniscule weights of $G$, of
which there are $\# Z$, where $Z$ is the center of $G$. 
 Moreover we gave a proof of a complete
list of vacua that are metastable, i.e. stable against the creation
of a single adjoint particle anti-particle pair.
These are the fundamental weights that are level 1 weights for affine
(non-twisted)  Kac-Moody algebras. This completes the partial 
proof in \cite{PSZ1}.

From the set-up it is clear that we might as well have introduced other
matter in a representation $S$, 
and that the same analysis would have boiled down to the 
following mathematical questions: for which representations $R$ is
 $C(R) \le C(R_i)$ for
all $R_i \subset R \otimes S$, and similarly,  for which representation
$R$ is  $C(R) \le C(R_i)$
 for
all $R_i \subset R \otimes S^n$
?  This question seems to be  of less physical importance,
since matter in a general representation
$S$  cannot be expected to mimick aspects of
 transverse gluons in higher dimensions, but the question
seems mathematically interesting, in view of the fact that it has such
 a non-trivial
answer for the adjoint representation. The techniques in this paper should
carry over, certainly, to the case of 
representations $R$ where the multiplicities of all
non-zero weights is 1.

Physically
 more interesting  is the question of stability of vacua in the
presence of adjoint matter and extra particles in a fundamental representation
of the gauge group. It should be easy to give the answer to this 
question algebraically, starting from the results obtained in this letter.
(There will be a similar parallel topological argument, 
based on the representation
of the center of the gauge group in the chosen fundamental representation.)

It would be very nice
to find a physical interpretation for the intriguing fact that
all metastable vacua are exactly the level 1 weights of affine (non-twisted)
Kac-Moody algebras (see also the suggestion in  \cite{PSZ1} of a phase
transition at finite mass).  
Moreover,
it seems interesting to generalize the explicit study of the different
vacua, and instanton contributions in 2d QCD (see e.g. \cite{BGV})
for different gauge groups. In fact, there seems to be a broad class of 
questions in two-dimensional gauge theory which have only been studied in
detail for $SU(N)$ gauge groups that should have interesting counterparts
for more general groups.

\section*{Acknowledgements}
\thanks{This work is supported in part by funds provided by the
U.S. Department of Energy (D.O.E.) under cooperative research
agreement DE-FC02-94ER40818.  }

\appendix

\section{A theorem}
\label{theorem}
We denote 
by $W$ the Weyl group, by $Q$ the root lattice, by $C$ the fundamental
chamber, and by $\rho$ half the sum of all positive roots. By $A$ we
denote the fundamental alcove and $\lambda_i$ are the fundamental weights,
corresponding to the simple roots $\alpha_i$.  $e_{\alpha_i}$ is the Lie
algebra generator associated with the simple root $\alpha_i$.

{\bf Theorem} (see e.g. \cite{OV}): 
The multiplicity of a representation $R(\nu)$ with highest
weight $\nu=\lambda+\mu'$ in the tensor product of
the representation $R(\lambda)$ with highest weight $\lambda$
and Dynkin labels $m_i=\lambda.\check{\alpha_i}$,
 and $R(\mu)$, where $\mu'$ is a weight
of $R(\mu)$, is equal to the dimension of the subspace of the
weightspace of $R(\mu)$ with weight $\mu'$ consisting of vectors
killed by $(e_{\alpha_i})^{m_i+1}$ for all simple roots $\alpha_i$. 

\section{ Miniscule and level 1 weights}
The miniscule weights are the non-zero dominant weights in the fundamental alcove.
They can be defined by $\mu.\check{\alpha_i} \ge 0$ and 
$\mu.\check{(\theta')} \le 1$. If we take, instead of $W \times Q$, the  group $W \times
Q'$ where $Q'$ is the {\bf Z}-span of the long roots, then the fundamental 
alcove becomes bigger: the inequality $\mu.\check{\theta}' \le 1$
is replaced by $\mu.\check{\theta}\le 1$,
 and intersects the dominant weights by all level 1 
weights of the affine (non-twisted) Kac-Moody algebra.
\label{Dynkin}
\begin{figure} 
 \epsfxsize=15cm
\epsfbox{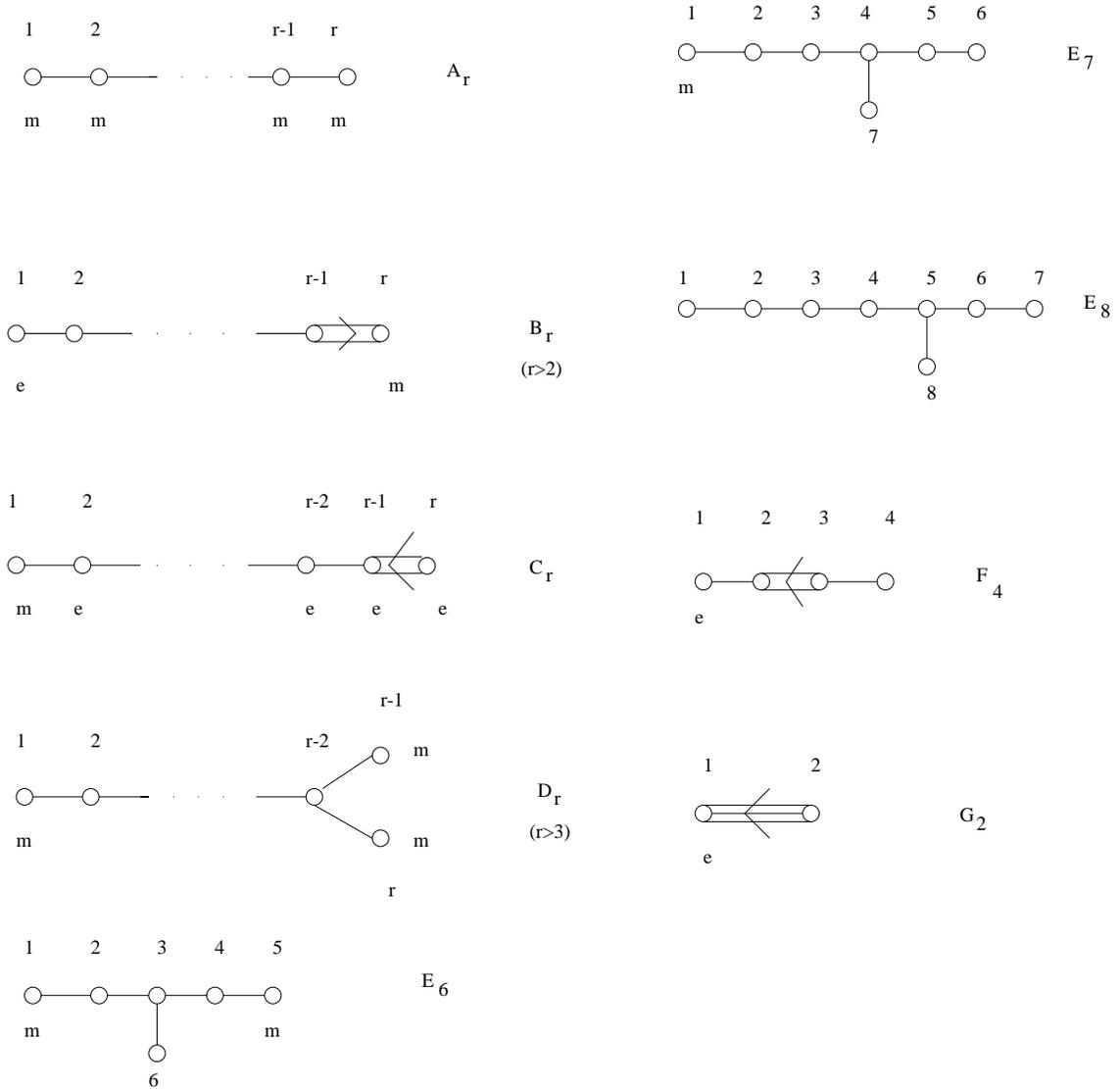}
\caption{\em We depict the Dynkin diagrams for all simple Lie algebras along
with a numbering of nodes,
 the non-zero miniscule weights (m), which are
all level 1,  as well as
the extra level 1 weights (e). All of them are fundamental weights and we
indicate the corresponding node in the Dynkin diagram. \em }
\end{figure}
\begin{table}
\begin{tabular}{|c||c|c|c|c|c|c|c|c|c||}
  \hline         
      & $A_r$ & $B_r$ & $C_r$ & $D_r$ & $E_6$ & $E_7$ & $E_8$ &
                            $F_4$ & $G_2$ \\ 
\hline
$\#$ level 1   & r+1 & 3 & r+1 & 4 & 3 & 2 & 1 & 2 & 2 \\
\hline
$\#$ miniscule & r+1 & 2 & 2 & 4 & 3 & 2 &1 & 1 &1 \\
\hline \hline
$\theta$       & $\lambda_1+\lambda_r$ & $\lambda_2$ & $2 \lambda_1$ &
$\lambda_2$ & $\lambda_6$ & $\lambda_6$  & $\lambda_1$ & $\lambda_4$ & $\lambda_2$ \\
\hline
\end{tabular}
\caption{\em We table the number of level 1 and miniscule weights for the finite simple
compact Lie algebras, and express their highest root $\theta$
 in terms of the fundamental
weights.}
\end{table}

  \end{document}